\documentclass[sigconf, letterpaper]{acmart}

\usepackage{rotating} 
\usepackage{subfigure}
\usepackage{adjustbox}
\usepackage{multirow}
\usepackage{multicol}
\usepackage{subfigure}
\usepackage{subcaption}
\usepackage[noabbrev]{cleveref}
\AtBeginDocument{%
  }

\begin{document}

\title{Towards Seamless Serverless Computing Across an Edge-Cloud Continuum}


\author{Emilian Simion}
\affiliation{%
  \institution{Institute of Informatics, University of Amsterdam}
  \institution{Faculty of Science, Vrije Universiteit Amsterdam}
  \streetaddress{}
  \city{Amsterdam}
  \country{The Netherlands}
}
\email{emilian.simion@student.uva.nl}

\author{Yuandou Wang}
\orcid{0000-0003-4694-9572}
\affiliation{%
	\institution{Multiscale Networked Systems, University of Amsterdam}
	\city{Amsterdam}
	\country{The Netherlands}}
\email{y.wang8@uva.nl}

\author{Hsiang-ling Tai}
\affiliation{%
  \institution{Informatics Institute, University of Amsterdam}
  \institution{Faculty of Science, Vrije Universiteit Amsterdam}
  \city{Amsterdam}
  \country{The Netherlands}}
\email{hsiang-ling.tai@student.uva.nl}

\author{Uraz Odyurt}
\orcid{0000-0003-1094-0234}
\affiliation{%
	\institution{High-Energy Physics, Radboud University}
	\city{Nijmegen}
	\country{The Netherlands}}
\affiliation{%
	\institution{Nikhef}
	\city{Amsterdam}
	\country{The Netherlands}}
\email{uodyurt@nikhef.nl}

\author{Zhiming Zhao}
\orcid{0000-0002-6717-9418}
\affiliation{%
	\institution{Multiscale Networked Systems, University of Amsterdam}
        \institution{LifeWatch ERIC Virtual Lab and Innovation Center (VLIC), Amsterdam, The Netherlands }
	\city{Amsterdam}
	\country{The Netherlands}}
\email{z.zhao@uva.nl}

\renewcommand{\shortauthors}{E. Simion et al.}

%

\begin{abstract}
Serverless computing has emerged as an attractive paradigm due to the efficiency of development and the ease of deployment without managing any underlying infrastructure. Nevertheless, serverless computing approaches face numerous challenges to unlock their full potential in hybrid environments. To gain a deeper understanding and firsthand knowledge of serverless computing in edge-cloud deployments, we review the current state of open-source serverless platforms and compare them based on predefined requirements. We then design and implement a serverless computing platform with a novel edge orchestration technique that seamlessly deploys serverless functions across the edge and cloud environments on top of the Knative serverless platform. Moreover, we propose an offloading strategy for edge environments and four different functions for experimentation and showcase the performance benefits of our solution. Our results demonstrate that such an approach can efficiently utilize both cloud and edge resources by dynamically offloading functions from the edge to the cloud during high activity, while reducing the overall application latency and increasing request throughput compared to an edge-only deployment.
\end{abstract}




\setcopyright{none}
\settopmatter{printacmref=false} 
\renewcommand\footnotetextcopyrightpermission[1]{} 
\pagestyle{plain}
\settopmatter{printfolios=true}
\maketitle

\section{Introduction}
Serverless computing, as a new and greatly popular paradigm with the cloud computing community, simplifies software development and deployment in a cloud environment. Serverless enables developers to run their code without having to manage any underlying infrastructure~\cite{baldini2017serverless, castro2019rise}. One typical offering is Functions-as-a-Services (FaaS), exemplified by public Cloud services such as AWS Lambda~\cite{sbarski2017serverless}, Azure Functions~\cite{kurniawan2019introduction}, and Google Cloud Functions~\cite{malawski2020serverless}. FaaS offers event-driven execution of functions for cloud customers to run their code for virtually any type of application or backend service without provisioning or managing any servers.

Apart from serverless computing, the emerging applications in various areas such as manufacturing~\cite{chen2018edge,qi2019smart}, healthcare~\cite{oueida2018edge,abdellatif2019edge}, smart cities~\cite{khan2020edge,lv2021intelligent}, agriculture and farming~\cite{o2019edge,hu2021blockchain} and transportation~\cite{zhou2021intelligent, lin2020edge}, have been both nurturing and demanding edge computing in recent years. Edge computing brings processing, data storage, and applications closer to the edge of the network, where end devices such as Internet-of-Things (IoT) devices and smartphones generate and consume data, which benefits from low latency, improved reliability, better data privacy, cost savings, and energy efficiency~\cite{ai2018edge, satyanarayanan2017emergence, 1036038}. 
At the same time, it is more common to have applications composed of different modules distributed over different tiers (e.g., edge, fog, cloud) and interoperate between themselves~\cite{morabito2023securing}. Such computational model has emerged under the term Edge-Cloud continuum, in which infrastructure's geo-distributed and heterogeneous nature presents unique challenges and opportunities~\cite{raith2022mobility, aslanpour2021serverless}. However, interoperating control across an Edge-Cloud continuum is still a challenge. Several existing works have made much progress in tackling this challenge. Nevertheless, resource management, scheduling, fault tolerance, deployment complexity, and cold-start mitigation~\cite{wang2021lass, tang2022distributed,gadepalli2019challenges, nastic2022serverless, ferry2022towards} still need to be addressed. 

In this paper, we address the interoperability problem between edge and cloud environments through serverless computing, making the deployment process more efficient across the Edge-Cloud continuum and improving the performance of processing applications. Our vision for interoperable serverless computing, which enables an Edge-Cloud continuum, involves a platform that abstracts not just the infrastructure, but also the location where functions are executed, using available resources as efficiently as possible. 
We aim to gain practical insights into the intricacies and challenges associated with this architecture. Our primary contribution can be summarized as follows: 
%
\begin{itemize}
    \item We review the current state of serverless platforms and compare them based on predefined requirements. 
    \item We implement a serverless platform and develop a novel edge orchestration technique that enables a seamless deployment of serverless functions across both edge and cloud environments. 
    \item We propose an edge offloading strategy and conduct extensive experiments to showcase its performance benefits. The source code is available at the GitHub repository\footnote{Knative Edge. \url{https://github.com/jevvk/knative-edge}}.
\end{itemize}

The remainder of this paper is structured as follows. \Cref{sec:related} presents related work and in \Cref{sec:ours}, we pose the requirement analysis, platform choices and our framework. \Cref{sec:experiments} details the experimental setup and discusses the results. Finally, we conclude our work in \Cref{sec:conclusion}. 

\section{Related Work}
\label{sec:related}
Recently, research has addressed serverless applications in the edge-cloud continuum, focusing on three dimensions: (1) scheduling functions in resource-limited edge environments~\cite{wang2021lass, tang2022distributed}, (2) optimizing serverless resource usage in edge environments~\cite{jeon2021deep, gadepalli2019challenges}, and (3) deployment complexity for seamless integration of serverless computing across this continuum~\cite{nastic2022serverless, ferry2022towards}. Wang \textit{et al.}~\cite{wang2021lass} introduced Lass, a platform for running latency-sensitive serverless apps on the edge using queuing theory to allocate resources and auto-scale as needed. Tang \textit{et al.}~\cite{tang2022distributed} proposed a deep learning task scheduling algorithm for resource utilization improvement at the edge. By contrast, we extend Knative's default round-robin scheduling to enable offloading requests from edge to cloud based on function response times. 
Gadepalli \textit{et al.}~\cite{gadepalli2019challenges} explored WebAssembly's potential for efficient serverless computing at the edge due to its low resource overhead. Jeon \textit{et al.}~\cite{jeon2021deep} optimized resource usage by caching function dependencies using deep reinforcement learning. Our approach focuses on offloading work to the cloud, not optimizing runtime overhead. Nastic \textit{et al.}~\cite{nastic2022serverless} presented Serverless Computing Fabric (SCF) for the Edge-Cloud continuum, addressing edge-native backend services, resource usage, and edge intelligence. 
Ferry \textit{et al.}~\cite{ferry2022towards} introduced a solution for Cloud-Edge-IoT applications with a modeling language. Our approach does not target IoT scenarios specifically; however, it may have higher overhead in some IoT contexts, as Knative's features have been designed with a focus on cloud environments.

\section{Methodology}
\label{sec:ours}
 
 

\subsection{Requirement Analysis}
The design of a serverless platform running on the Edge-Cloud continuum necessitates careful consideration of multiple factors, including scalability, security, cost-effectiveness, and performance. On the one hand, this method must be capable of operating at both the Edge, close to the source of data, as well as in the Cloud, where it can leverage the resources of a large data center. On the other hand, the design should address the unique challenges presented by this hybrid environment, such as managing data transfer between the Edge and Cloud and handling variable traffic levels in real-time. By harnessing the advantages of both the Edge and the Cloud, our serverless platform can provide a flexible and efficient solution for a wide range of applications. To achieve this, we define the following functional and non-functional requirements. 

The \textbf{functional requirements} we propose for our solution are the following:
\begin{itemize}
    \item \textbf{Location Agnostic}. The same function definition can be utilized to run serverless functions in both Edge and Cloud environments.
    \item \textbf{Scalability}. The platform is capable of accommodating the addition of new edge clusters and nodes dynamically.
    \item \textbf{Dynamic Scheduling}. The edge cluster gateway must have the ability to dynamically determine the execution location of a function, either at the Edge or in the Cloud.
\end{itemize}

We also define the \textbf{non-functional requirements} as follows:
\begin{itemize}
    \item \textbf{Heterogeneity}. The system should support nodes with varying degrees of hardware capabilities (e.g. x86, ARM).
    \item \textbf{Resource Allocation}. The system must possess the capability to dynamically allocate resources in accordance with workload demands.
    \item \textbf{Fault Tolerance}. If a worker node experiences a failure, the system can still operate normally.
    \item \textbf{Reliability}. The system should be robust to bad Edge-Cloud connection.
    \item \textbf{Security}. Communication between Edge and Cloud should be secure.
\end{itemize}

We establish a set of \textbf{selection criteria} for platform comparison that will guide the design process. 
\begin{enumerate}
    \item The serverless platform should be actively maintained in the community.
    \item The serverless platform should be open-source and have a permissive license. We wish to be able to extend the platform while avoiding any restrictive licensing.
    \item The serverless platform must possess the capability to scale-to-zero and scale workloads according to actual demand, which are essential features for its intended purpose.
    \item The serverless platform is able to run on limited resources and a varying degree of heterogeneous hardware.
\end{enumerate}
These criteria serve as the foundation for our solution, ensuring that the final solution aligns with the desired goals and objectives.

\subsection{Technology Investigation}
\begin{table}[!th]
    \centering
    \vspace{-3mm}
    \caption{For each investigated serverless platform, we have assessed whether they fit our platform criteria \#(1-4).}\label{tab:platform-critera}
    \begin{tabular}{lcccc}
    \toprule
        \multirow{2}*{\textbf{Serverless Platform}} & \multicolumn{4}{c}{\textbf{Platform criteria}} \\ \cline{2-5} 
        \textbf{} & \textbf{\#(1)} & \textbf{\#(2)} & \textbf{\#(3)} & \textbf{\#(4)} \\
    \midrule
    Kyma        & $\checkmark$   & $\checkmark$   & $\checkmark$   & $\checkmark$   \\ 
    Knative     & $\checkmark$   & $\checkmark$   & $\checkmark$   & $\checkmark$   \\ 
    OpenFaaS    & $\checkmark$   & {}             & $\checkmark$   & $\checkmark$   \\ 
    Fission     & $\checkmark$   & $\checkmark$   & $\checkmark$   & $\checkmark$   \\ 
    OpenLambda  & {}             & $\checkmark$   & $\checkmark$   & $\checkmark$   \\ 
    OpenWhisk   & $\checkmark$   & $\checkmark$   & $\checkmark$   & $\checkmark$   \\ 
    Kubeless    & {}             & $\checkmark$   & $\checkmark$   & $\checkmark$   \\ 
    Fn          & {}             & $\checkmark$   & $\checkmark$   & $\checkmark$   \\ 
    IronFunctions   & {}         & $\checkmark$   & $\checkmark$   & $\checkmark$   \\
    \bottomrule
    \end{tabular}
    \vspace{-3mm}
\end{table}
We collect information from literature and official sources, such as GitHub statistics incl., stars, forks, issues, number of commits, and software documentation, and conclude nine mainstream open-source serverless platforms.
We have checked the capabilities of each platform as understood from the official documentation and initial deployments of the platforms on our experimental setup and evaluated the reasons for including or excluding them from our platform options. This evaluation process helps us to determine which serverless platforms are best suited to meet our platform's functional and non-functional requirements and to make informed decisions about which platform to choose for our implementation. We summarize our findings in \Cref{tab:platform-critera} and explain our reasoning below for each serverless platform on a case-by-case basis.
\begin{itemize}
    \item \textbf{Knative} is a well-regarded serverless platform that has gained popularity in academic research circles and within the open-source community. Our requirements analysis has determined that it is a suitable choice to include in the platform options.
    \item \textbf{Kyma}. As a FaaS solution based on \verb|Knative|, \verb|Kyma| inherits this underlying framework's technological capabilities and constraints. It meets all the previously defined requirements.
    \item \textbf{OpenFaaS}. Despite its technical capabilities, the licensing of core serverless features in \verb|OpenFaaS| prevents us from using it; it requires a paid license for crucial features such as scaling to zero and event handling using message brokers.
    \item \textbf{Fission}. Our analysis has indicated that it has the capabilities and features to fulfill the intended purpose of the platform, as well as meet the non-functional requirements such as performance, scalability, security, and cost-effectiveness.
    \item \textbf{OpenLambda}. Despite being designed for academic research, \verb|OpenLambda| lacks more support for cluster deployments, which is a critical requirement for deploying our solution.
    \item \textbf{OpenWhisk} is a well-established serverless platform. Our requirements analysis indicates that it satisfies all the previously established functional and non-functional requirements. 
    \item \textbf{Kubeless}. Despite being widely recognized for its performance, we exclude \verb|Kubeless| from the design of the serverless platform because it is no longer actively maintained.
    \item \textbf{Fn}. While \verb|Fn| was one of the pioneers in the open-source serverless space, it is no longer maintained and will not be considered in our analysis.
    \item \textbf{IronFunctions}. Like \verb|Fn|, \verb|IronFunctions| is an early example of an open-source serverless solution. Unfortunately, like \verb|Fn| and \verb|Kubeless|, it has become inactive and, as a result, has been excluded from our design's list of platform options. 
\end{itemize}

During our practical assessment using our test bed, we encountered various deployment challenges for different serverless platforms. Notably, we faced memory limitations that prevented us from deploying \verb|OpenWhisk| due to requirements associated with the necessary message broker. While \verb|Fission| was successfully deployed (as shown in \Cref{tab:platform-critera}), it exhibited suboptimal performance on our test bed, with function instances frequently hanging and necessitating frequent restarts. In contrast, both \verb|Knative| and \verb|Kyma| were successfully deployed and thoroughly tested. After careful consideration, we selected \verb|Knative| as the foundation for our solution. This choice was supported by the fact that \verb|Kyma| is built on top of \verb|Knative|, ensuring compatibility between our solution and \verb|Kyma|.

\subsection{Platform Design and Implementation}
\begin{figure*}[!t]
    \centering
    \includegraphics[width=\textwidth]{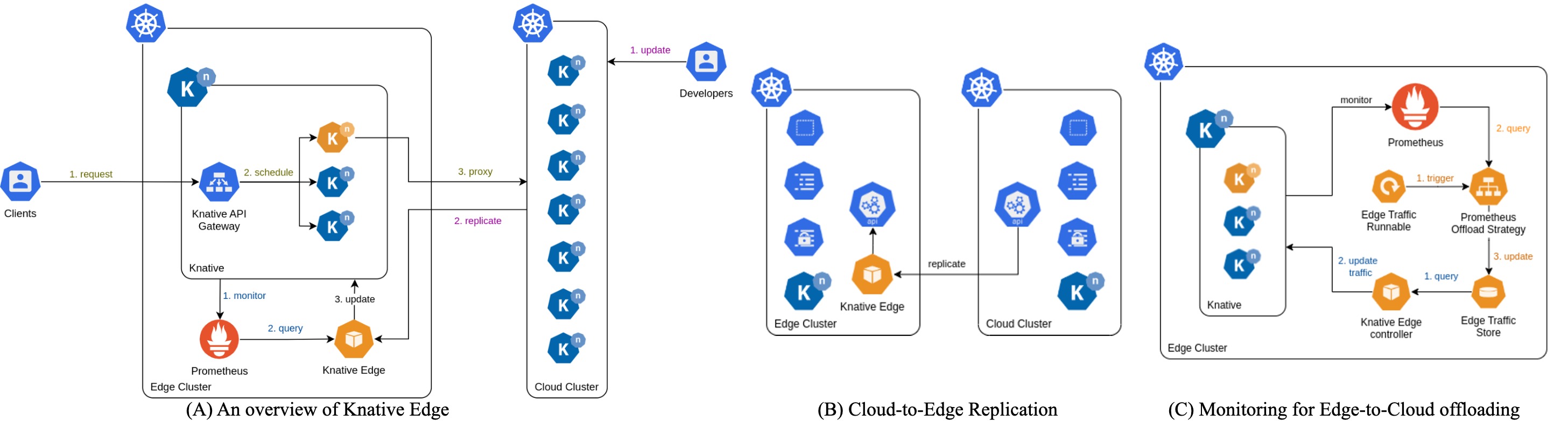}
    \caption{An overview of the System Architecture and core components.}
    \label{fig:architecture}
\end{figure*}
We design and implement \verb|Knative Edge| as an extension of the existing \verb|Knative| serverless platform as shown in \Cref{fig:architecture}(A). 



\subsubsection{Cloud-to-Edge Replication}
The \verb|Knative Edge| controller mirrors \verb|Knative Services| from the Cloud cluster to the Edge cluster by watching for changes in the Kubernetes resources of both clusters and modifying the Edge resource to keep a consistent definition. As shown in the \verb|replicate| of \Cref{fig:architecture}(B), it uses several different components within the edge cluster to achieve its goals.
However, replicating a resource may risk making unnecessary changes to the resources it replicates. It is most evident when a \verb|Knative Service| is replicated, as any changes can trigger \verb|Knative Serving| to react and make more changes to the resource, which can trigger \verb|Knative Edge| to make more changes. This feedback loop can cause degradation of the serverless functions running either on the Edge or Cloud and can increase the traffic between the Cloud and Edge.

Our approach to this issue is to selectively compare fields in the \verb|Knative Service| definition. Whenever \verb|Knative Edge| receives an update for services it replicates, it copies the current definition from the Edge and overwrites a subset of its fields using the definition from the Cloud. When overwriting, we skip over the state of the service and any annotations which are not defined by \verb|Knative Edge|; thus, the internal state of the Edge resources are persisted. The new service definition is compared to the current definition on the Edge cluster and deployed to the cluster if any change is detected.

\subsubsection{Edge-to-Cloud Offloading}
For efficiently monitoring and managing runtime platform metrics in our Edge cluster, we utilize an instance of Prometheus deployed on each Edge cluster, as shown in \Cref{fig:architecture} (C). This instance is configured to be aware of all \verb|Knative| components in the cluster, including all running function instances. It scrapes these metrics regularly and temporarily stores them in a time series database, which is optimized for storing and querying metrics such as those generated by \verb|Knative|. Since only recent data is being queried by \verb|Knative Edge|, we configure short data liveness to reduce as much as possible the overhead of \verb|Prometheus|. 
The scheduling is driven by a load-balancing algorithm that spreads the traffic to the different routes based on a defined percentage. We implement a simple default offloading strategy that uses the request latency metrics of all the functions running at the Edge. The API Gateway makes the decision randomly, and only a percentage of traffic (decided by the offloading strategy) is being sent to the cloud.

Let $\mathcal{X}_{l}(t)$ be the distribution of request latencies at time $t$ and $p_{95}$ and $p_{50}$ are the $95^{th}$ and $50^{th}$ percentile, respectively. The weighted sum of the latest latency response ratio $r_{l}(t)$ can be given by,
\begin{equation}\label{eq:rlatency}
    r_{l}(t) = \frac{p_{95}(\mathcal{X}_{l}(t))}{p_{50}(\mathcal{X}_{l}(t))}
\end{equation}
Giving more importance to recent values than to older ones, we use an exponentially decreasing weighted sum with $c_{\text{decay}}$ as exponent to implement $r^{'}_{l}(t)$, which is given by, 
\begin{equation}\label{eq:r'latency}
    r^{'}_{l}(t) = \frac{\sum_{k=0}^{c_t}c^{k}_{\text{decay}} \times r_{l}(t-k)}{\sum^{c_t}_{k=0}c^k_d}
\end{equation}
where $c_t$ defines the how many steps in time are used to calculate the weighted sum of the latest request latency ratios measured. It calculates $r_{t}$ in \Cref{eq:rtraffic}, the intended traffic percentage that should be forwarded to the Cloud. 
Through different iterations, we found that directly using $r_{t}$ for setting the traffic percentage lead to unstable offloading. We define $r_{t}(t)$ as
\begin{align}\label{eq:rtraffic}
    r_{t}(t) = \begin{cases}
        0, \text{\quad if } r^{'}_{l}(t)< c_\text{soft},\\
        100, \text{\quad if } r^{'}_{l}(t) > c_{\text{hard}}, \\
        100 \times \frac{r^{'}_{l}(t) - c_{\text{soft}}}{c_{\text{hard}} - c_{\text{soft}}}, \text{\quad otherwise}.
    \end{cases}
\end{align}
in which $c_{\text{soft}}$ and $c_{\text{hard}}$ are the soft limit and hard limit of the percentile ratio. For any $r^{'}_{l}$ that falls bellow $c_{\text{soft}}$, the traffic percentage is set to 0. For values above $c_{\text{hard}}$, the traffic percentage is set to 100. And finally, for values between, we interpolate them between 0 and 100, depending on where the values lie in respect to $c_{\text{soft}}$ and $c_{\text{hard}}$. Hence, we define $R_{t}$ in \Cref{eq:Rtraffic} as a more stable way of updating the traffic percentage.

\begin{align}\label{eq:Rtraffic}
    R_{t}(t) &= R_{t}(t-1)\times c_{\text{in}}+r_{t}(t)\times (1-c_{\text{in}}), R_{t}(0)= 0.
\end{align}
where $c_{\text{in}}$ is the inertia factor, which measures how much $r_{t}$ influences $R_{t}$.



\section{Platform Evaluation}
\label{sec:experiments}

\begin{table}[!ht]
    \centering
    \caption{For each workload type and traffic split, we present the total number of successful responses.}
    \label{tab:experiment-requests-served}
    \begin{tabular}{ccccc}
        \toprule
         \textbf{Traffic} & {\textbf{MatMult}} & {\textbf{Image Proc.}} & {\textbf{ I/O}} & {\textbf{Mixed} } \\
        \midrule
        $0\%$   &$2406$  & $3627$ & $4852$ & $4152$\\
        $25\%$  & $2699$ & $4044$ &$5947$ & $5237$ \\
        $50\%$  & $2664$ & $4045$ &$9371$ & $7486$  \\
        $75\%$  & $2683$ & $3970$ & $9371$ &$8619$ \\
        $100\%$ & $2668$ &$3969$ & $9408$ &$8725$ \\
        $auto$  &$2700$  & $4016$ & $6548$ & $7989$ \\
        \bottomrule
    \end{tabular}
    \vspace{-4mm}
\end{table}
\begin{figure*}[!th]
    \centering
    \begin{minipage}[b]{0.495\textwidth}
        \centering
        \includegraphics[width=\textwidth]{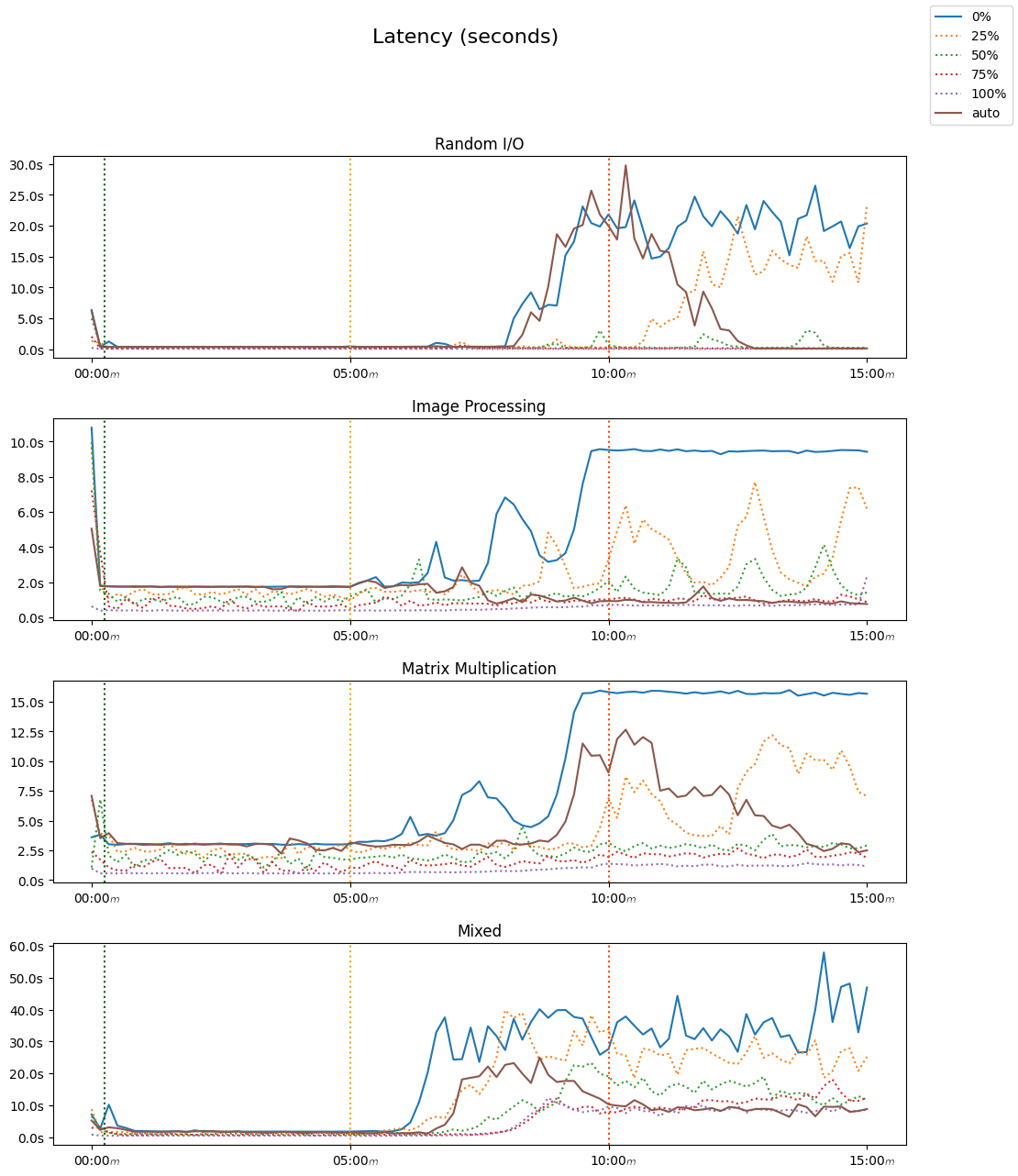}
        \label{fig: latency}
    \end{minipage}
    \hfill
    \begin{minipage}[b]{0.495\textwidth}
        \centering
        \includegraphics[width=\textwidth]{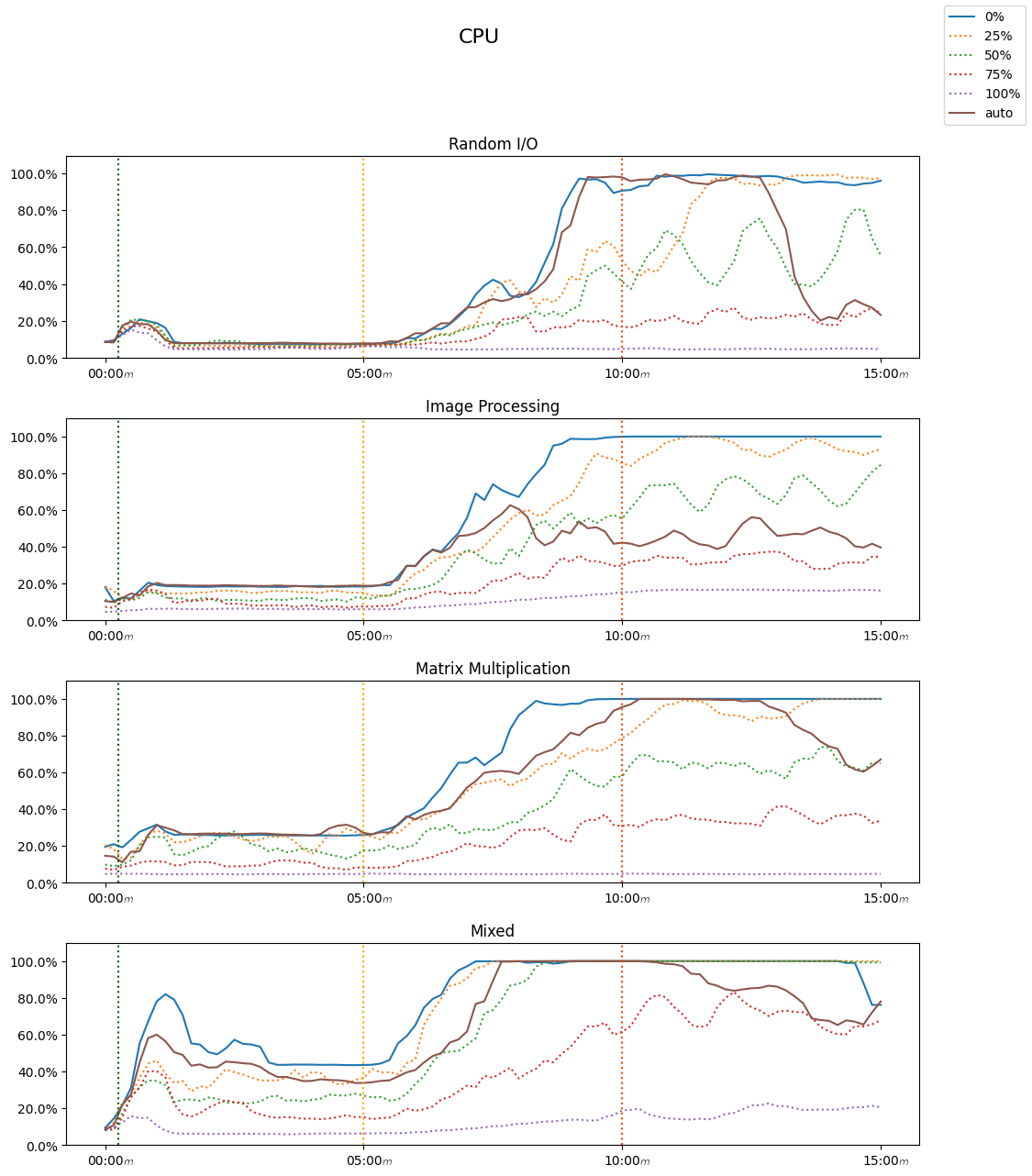}
        \label{fig: cpu}
    \end{minipage}
    \hfill
    \begin{minipage}[b]{0.495\textwidth}
        \includegraphics[width=\textwidth]{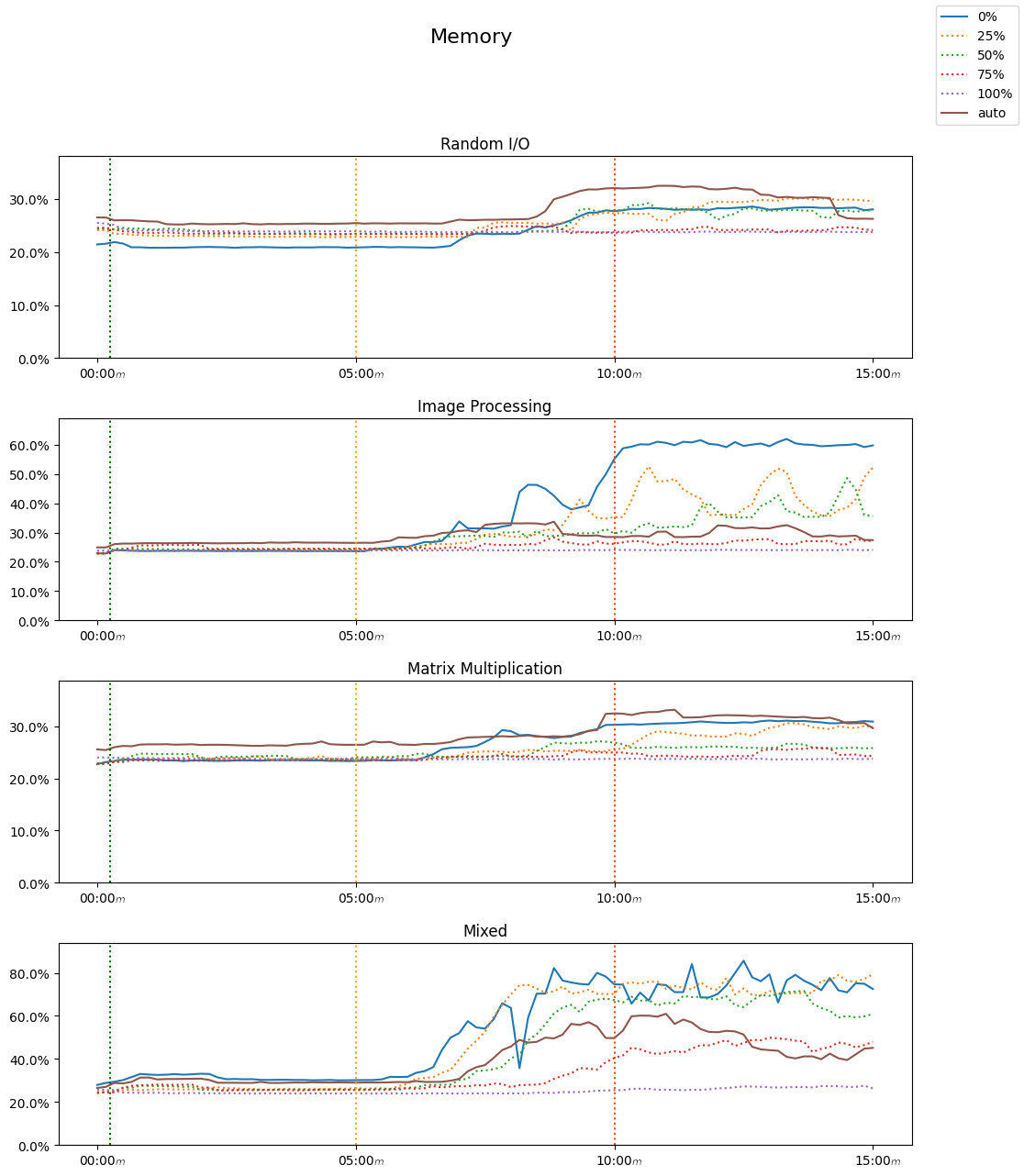}
        \label{fig: mem}
    \end{minipage}
    \hfill
    \begin{minipage}[b]{0.498\textwidth}
        \centering
        \includegraphics[width=\textwidth]{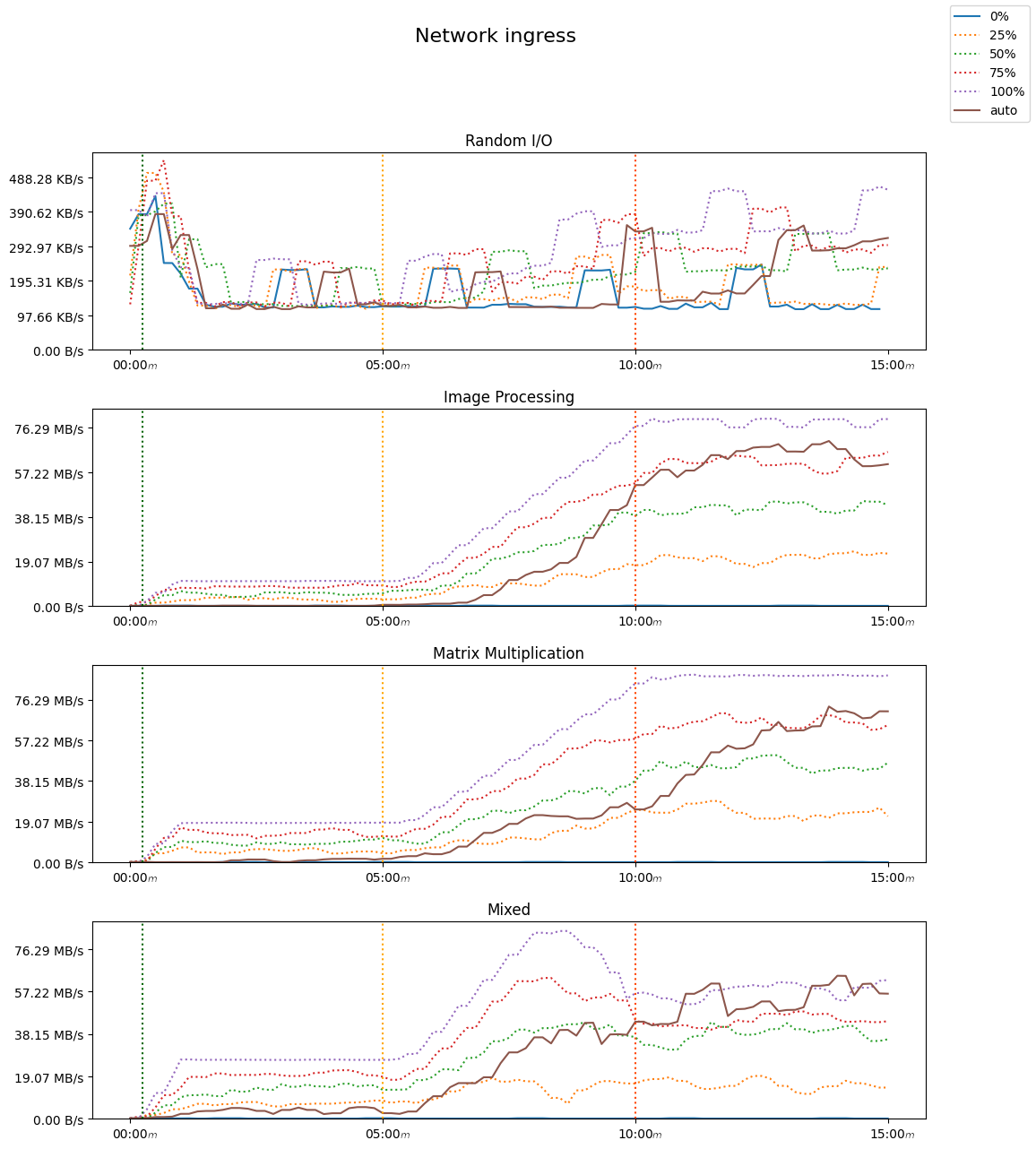}
    \label{fig: network-ingress}
    \end{minipage}
    \caption{Performance results, including latency (seconds), CPU, memory, and network usage, are presented for workload traffics (refer to \Cref{tab:experiment-requests-served}). }
    \label{fig: perf}
    \vspace{-4mm}
\end{figure*}

\subsection{Experimental Setup}
We utilized four Raspberry Pi 3B+ devices, a low-power x64 edge device, and a cloud virtual machine (VM) to establish our platform. This platform was meticulously developed and rigorously tested on Knative Serving 1.7 and Kubernetes 1.24. Our experiments involved four real-world workloads: matrix multiplication (MatMult), image processing (Image Proc.), random I/O, and a combination of these three loads (Mixed). These workloads exhibit varying demands on CPU, memory, disk, and network resources during execution, making them suitable targets for evaluating our platform's capabilities. We generated requests for these workloads using a specialized experiment runner, allowing us to control the request rate: 
initially, we used a low request rate, then increased the rate linearly to a high request rate, which was maintained until the end of the run. The rates were chosen in such a way that the low request rate could be handled entirely by the edge devices and the high request rate would overload the edge devices under no offloading.
Additionally, we employed an edge scheduling strategy to determine how traffic was distributed between the Edge and the Cloud. This strategy could be configured to various distribution levels (0\%, 25\%, 50\%, 75\%, 100\%, and auto) in our experiments, as detailed in \Cref{tab:experiment-requests-served}. These settings enabled us to assess the impact of different workload types and edge scheduling strategies on the platform's performance. 
We measured several performance metrics, including average response time, CPU and memory utilization of the edge devices, as well as network utilization of the cloud VM.

\subsection{Results and Evaluation}\label{sec:results}
\Cref{fig: perf} illustrates the average latency of the responses, the average CPU and memory utilization of the edge devices, and network utilization of the cloud VM, all measured over the course of each experiment run. 

\textbf{Latency.}
The offloading functionality improves considerably response times. For instance, our solution initially exhibits slower reaction times, leading to reduced response times. However, as more requests are offloaded to the Cloud, the response time converges towards the lower bound given by our edge network conditions as soon as the request rate is increased (see \Cref{fig: perf} (Latency)).

\textbf{CPU}. 
Our results indicate that our solution can effectively offload excessive workload from the Edge cluster but may also underserve many requests (see \Cref{fig: perf} (CPU)). As our algorithm optimizes the response times of workloads, CPU utilization is reduced as a consequence. However, determining the optimal level of CPU utilization is more complex. If the CPUs of edge devices are not fully utilized but requests are being forwarded to the Cloud, Edge cannot be efficiently utilized. Furthermore, reserving spare CPU resources could also benefit the Edge cluster when they are needed for possible bursts of requests or system operations. Finally, our test bed uses Raspberry Pi 3B+ as edge devices which have a low CPU power, and more powerful edge devices, such as Nvidia Jetson, might need more analysis and investigation in order to optimally balance where requests should be served from.

\textbf{Memory}. 
We assess that, in most cases, memory utilization remains within reasonable bounds. Our initial assessment of the workloads suggested that we would observe a significant increase in memory utilization for both image processing and matrix multiplication. While the latter did see a noticeable increase in memory consumption, the former has seen a much-lessened effect, performing similarly to the random I/O workload, which we estimated to have no increase in memory (see \Cref{fig: perf} (Memory)).


\textbf{Network}. 
Bandwidth saturates for image processing and matrix multiplication when all requests offload to the cloud, however the mixed workload never hits the maximum of 100MB/s. Similarly, our network ingress findings indicate that image processing and matrix multiplication are constrained by Edge-to-Cloud network bandwidth at full offloading. In case the network is the bottleneck (which is in the case of the matrix multiply and mixed workloads), then the offloading does not help. It would make the response times worse, depending on how they compare to the edge. Additionally, our offloading strategy does not take into account the network latency or bandwidth between the Edge and Cloud environments, and a more sophisticated strategy is required to optimally offload in different network conditions.

\section{Conclusion}
\label{sec:conclusion}
The topic of serverless computing on the Edge-Cloud continuum is still in its infancy. This paper aims to provide valuable insights into the design of serverless platforms for a unified Edge-Cloud computing environment. We presented our platform, an extension of the serverless platform Knative that enables deployments across edge and cloud environments and offloads requests from edge devices to the cloud. We demonstrated that our approach can deploy a serverless application across edge and cloud environments and demonstrated its offloading capability. In future, we will explore more about offloading strategies for resource optimization and performance improvement.

\begin{acks}
This work has been partially funded by the European Union's Horizon 2020 research and innovation program through the project CLARIFY (860627), the ENVRI-FAIR (824068) project, BlueCloud-2026 (101094227) project, by the NWO LITE-LIFE project and by the LifeWatch ERIC.
\end{acks}

\bibliographystyle{acm}
\bibliography{refs}

\end{document}